\documentclass[aps,pra,twocolumn,groupedaddress]{revtex4}

\usepackage[]{natbib}
\usepackage[]{graphicx}
\usepackage{amssymb}

\newcommand{\pic}[2]{\includegraphics[width=#1\textwidth]{#2}}

\begin{document}

\title{Lattice Thermodynamics for Ultra-Cold Atoms}

\author{D. McKay}
\author{M. White}
\author{B. DeMarco}
\email[email:]{bdemarco@illinois.edu}
\affiliation{Department of Physics, University of Illinois at Urbana-Champaign, Urbana, Illinois 61801, USA}

\date{\today}

\begin{abstract}
We measure the temperature of ultra-cold $^{87}$Rb gases transferred into an optical lattice and compare to non-interacting thermodynamics for a combined lattice--parabolic potential.  Absolute temperature is determined at low temperature by fitting quasimomentum distributions obtained using bandmapping, i.e., turning off the lattice potential slowly compared with the bandgap.  We show that distributions obtained at high temperature employing this technique are not quasimomentum distributions through numerical simulations.  To overcome this limitation, we extract temperature using the in-trap size of the gas.  
\end{abstract}


\maketitle

Ultra-cold atoms confined in optical lattices are a promising system for studying models of strongly correlated systems relevant to condensed matter physics.  Experiments have been able to observe a quantum phase transition from a superfluid to Mott-insulating state for bosons \cite{greiner:2002,spielman:2008,kohl:2005}, the superfluid-to-insulator transition for bosons \cite{mun:2007}, super-exchange \cite{trotzky:2008}, the cross-over between quantum tunneling and thermal activation of phase slips \cite{mckay:2008}, reversible depletion of condensate fraction induced by fine-grained disorder \cite{white:2008}, and evidence for a Mott insulator of fermions \cite{schneider:2008,jordens:2008}.  Straightforward interpretation of these results in certain cases has been complicated by difficulties in measuring temperature related to strong interactions and the lattice potential \cite{diener:2007}. 

Temperature has been measured for both bosonic \cite{anderson:1995} and fermionic \cite{demarco:1999} atom gases confined in harmonic traps by fitting the momentum distribution, obtained after ballistic expansion, to analytic expressions obtained using the semi-classical approximation \cite{bagnato:1987,butts:1997}. Unfortunately, this approximation fails in an optical lattice because of the rapid spatial variations in the potential \cite{butts:1997,pethick:2001}.  Despite this limitation, progress has been made on measuring temperature in an optical lattice using alternative techniques. One method involves site occupancy statistics \cite{stoferle:2006}, which can be straightforwardly related to temperature in the atomic (i.e., tunneling energy much less than interaction energy) limit \cite{kohl:2006}. Another technique is to determine temperature using the visibility of the momentum distribution \cite{gerbier:2005}. This method, however, has generated some controversy \cite{kato:2008,pollet:2008,diener:2007,gerbier:2007b}, cannot be used at high temperature, loses sensitivity at very low temperature, and does not work for fermions.  Finally, advances have been made employing quasimomentum distributions to measure temperature. By fitting to quasimomentum distributions determined from a restricted region of a momentum profile, the condensate fraction and temperature have been measured for a lattice in the limit described by the 2D Bose-Hubbard model \cite{spielman:2008}.  These experiments, however, were carried out at a fixed temperature below the critical temperature $T_c$ for Bose condensation.  In this work, we extend this technique over a wide range of temperatures to an ultra-cold gas of non-condensed $^{87}$Rb atoms confined in a 3D optical lattice. 

In contrast to \cite{spielman:2008}, we obtain quasimomentum distributions \cite{yi:2007,lin:2008,kohl:2005b} through bandmapping \cite{greiner:2001,denschlag:2002}, a procedure in which the lattice potential is turned off slowly compared with the bandgap before time-of-flight imaging.  We fit these distributions to a semi-classical analysis to determine temperature.  To explore a range of temperatures in the lattice we vary the temperature of the atoms before transfer into the lattice, when they are confined in a purely harmonic potential.  The temperature determined using this method is compared with thermodynamic predictions for non-interacting particles, assuming adiabatic transfer from the parabolic potential.  We show that this procedure is successful at low temperatures, but fails at sufficiently high temperature such that states with high quasimomentum and localized states are occupied.  Through numerical simulation, we show that this failure is related to a breakdown of bandmapping in this regime.  At high temperature we demonstrate that temperature can alternatively be extracted using the in-trap density distribution. 

The paper is organized as follows: Section \ref{sect1} reviews solutions to the Schr\"{o}dinger equation for a combined lattice--harmonic potential; the single particle eigenstates are applied to calculate the lattice thermodynamics and thereby predict temperature after adiabatic transfer from a purely harmonic potential.  In Section \ref{sect2} we discuss the quasimomentum distribution of a thermal gas and imaging of that distribution using our apparatus.  In Section \ref{sect3} we show measurements of temperature for a thermal gas obtained using quasimomentum distributions.  In Section \ref{sect4} we present a numerical simulation of bandmapping to explain the breakdown of this method at high temperatures. In Section \ref{sect5} we will discuss determining temperature using in-trap size and present experimental results. 

\section{Thermodynamics of the Combined Lattice--Parabolic Potential \label{sect1}}

In most optical lattice experiments to date, an ultra-cold gas of atoms is first created in a parabolic potential, followed by a slow turn-on of the periodic lattice potential.  The temperature in the final combined lattice--parabolic potential can be predicted by measuring the temperature in the harmonic potential, calculating the corresponding entropy, and assuming that the turn-on is isentropic.  For this procedure, the entropy in the lattice--parabolic potential must be calculated, which has been done for non-interacting particles \cite{blakie:2004,blakie:2007b,blakie:2005,blakie:2007,kohl:2006}, in the mean-field approximation \cite{yoshimura:2008}, and for the strongly interacting limit\cite{rey:2006,ho:2007,cramer:2008,gerbier:2007}. In the general case, a quantum Monte Carlo calculation is necessary \cite{pollet:2008}. 

We employ a simple but exact non-interacting theory to predict temperature in the lattice--parabolic potential.  In this section, we review the eigenstates for this potential, which were reported in Ref. \cite{rey:2005}.   We briefly discuss how these states are used to calculate the entropy in the lattice, and thereby to predict the temperature and condensate fraction after loading from a purely harmonic trap. 

The single-particle Hamiltonian describing this system is:
\begin{eqnarray}
H & = & \sum_{i=x,y,z} \left\{\frac{p_i^2}{2m}+ \frac{s E_{R}}{2}\left[1-\cos(\pi x_i/d)\right] \right. \nonumber \\*
& & \left.  + \frac{1}{2}m\omega_i^2 x_i^2 \right\}, \label{eqn:ham1}
\end{eqnarray}
where $sE_R$ is the lattice depth  $(E_R=(h\pi/d)^2/2m$ is the recoil energy of the atom), $m$ is the mass of the atom, $d$ is the lattice spacing, and $\omega$ is the external harmonic confinement frequency.  Some intuition about the nature of the single-particle eigenstates can be gained by considering the case when the harmonic potential is absent.  When $\omega=0$, the solutions to this Hamiltonian are Bloch wavefunctions,
\begin{equation}
\Phi_{\vec{q},l}(\vec{x}) = e^{i \vec{q}\cdot \vec{x}/\hbar} u_{\vec{q},l} (\vec{x}),
\end{equation}
where $\vec{q}$ is the quasimomentum of the state, $l$ specifies the band-index, and $u_{\vec{q},l} (\vec{x})$ is a function with the same periodicity as the lattice.  An appropriate sum over Bloch wavefunctions defines the Wannier function centered at site $j$ \cite{ashcroft:1976}, 
\begin{equation}
w_{j,l}(\vec{x}) = \int d^3q~e^{i \vec{q} \cdot \vec{R}_j/\hbar} \Phi_{\vec{q},l} (\vec{x}).
\end{equation}

Experiments with ultra-cold atoms are typically confined to the lowest band, so we drop the band index in the discussion that follows and let $l=0$.  The Hamiltonian can also be conveniently expressed in second-quantized notation, in which the operator $\hat{a}_i^{\dagger}$ creates a particle with wavefunction $w_i(\vec{x})$. In the tight-binding approximation (accurate for $s \gtrsim 4$), Eq. \ref{eqn:ham1} becomes:
\begin{equation}
H = -J \sum_{<i,j>} \hat{a}_i^{\dagger} \hat{a}_j + \frac{1}{2}m\omega^2 \sum_i r_i^2 \hat{n}_i, \label{eqn:ham2}
\end{equation}
where \cite{jaksch:1998},
\begin{eqnarray}
J & = & \int d^3x~w^{*}_i(\vec{x})\left[-\frac{\hbar^2}{2m}\nabla^2 \right. \nonumber \\*
& & \left. + \frac{s E_{R}}{2}\left(1-\cos(\pi \vec{x}/d)\right)\right]w_{i+1}(\vec{x})
\end{eqnarray}
($J=\frac{4}{\sqrt{\pi}} E_R s^{3/4} e^{-2\sqrt{s}}$ in the limit $V_0 \gg E_{R}$ \cite{zwerger:2003}) is the nearest-neighbor tunneling matrix element, $i$ and $j$ index lattice sites, $\langle \rangle$ represents a sum over nearest-neighbors, and $\hat{n}_i = \hat{a}^{\dagger}_i \hat{a}_i$ is the number of atoms at site $i$.  For $\omega=0$, the energy spectrum is given by the usual tight-binding dispersion relation,
\begin{equation}
E = \left\langle \Phi_{\vec{q}}\right| H \left| \Phi_{\vec{q}} \right\rangle = 2J \sum_{i=x,y,z} \left[1-\cos \left(\frac{\pi}{q_B} q_i\right) \right], \label{eqn:dispersion}
\end{equation}
where $q$ is the quasimomentum, and $|q|<q_B=\hbar \pi/d$ ($q_B$ is the Brillouin-zone momentum).  The dispersion relation is characterized by an energy gap at $q_B$ between the lowest energy band and the first excited band. 

The parabolic potential present in ultra-cold atom lattice experiments breaks the periodic symmetry of the lattice potential and therefore changes the single-particle eigenstates \cite{hooley:2004}.  Recently, an analytic solution was derived for the eigenstates and eigenenergies for the combined parabolic--lattice potential \cite{rey:2005}, which enables straightforward calculation of non-interacting thermodynamics.  From these solutions, the energy of a state with quantum numbers $n_x,n_y,n_z$ for the Hamiltonian in Eq. \ref{eqn:ham1} is $E_{n_x,n_y,n_z}=E_{n_x}+E_{n_y}+E_{n_z}$ \cite{reynote} with
\begin{equation}
E_{n} = \left\{ \begin{array} {lr} \frac{\Omega}{4} a_{n}(\alpha), & \mbox{$n$ even} \\ \frac{\Omega}{4} b_{n+1}(\alpha), & \mbox{$n$ odd} \end{array} \right., \label{eqn:eigenergies}
\end{equation}
where $n=0,1,2,3,\ldots$ is an integer, $\Omega= m \omega^2 d^2/2$, $\alpha=4J/\Omega$, and $a_{n} (\alpha)$ and $b_{n+1} (\alpha)$ are the Mathieu characteristic values.

\begin{figure}
 \pic{0.5}{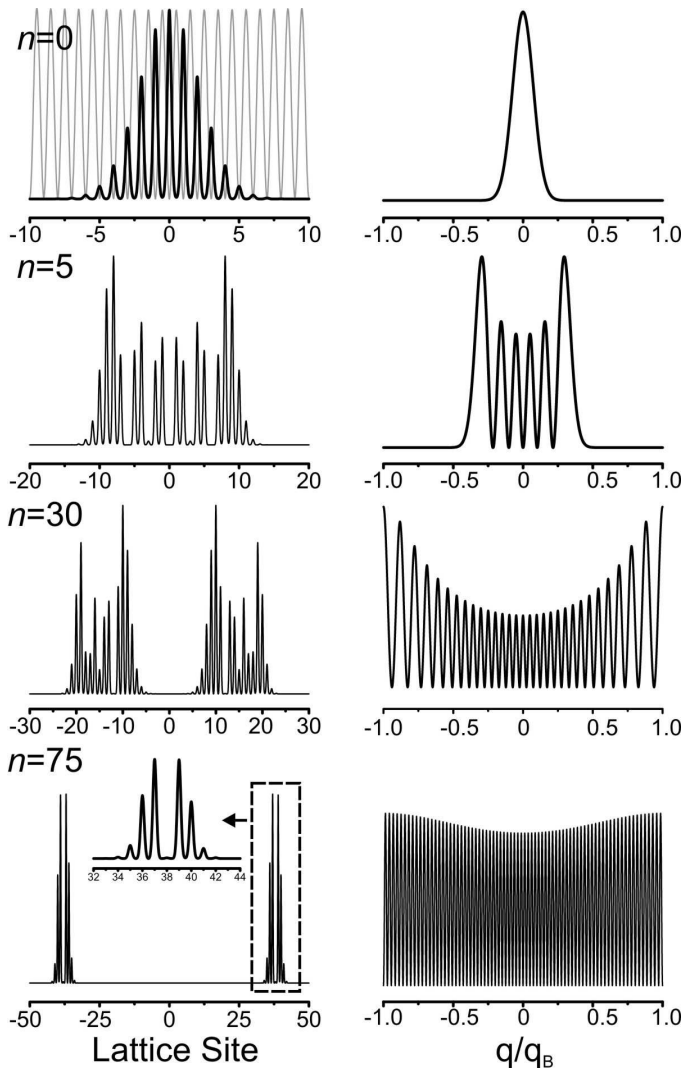}
 \caption{Spatial (left column) and quasimomentum (right column) probability distributions for a one dimensional combined lattice-parabolic potential (light grey curve in top left image) for different quantum numbers $n$ and $\alpha=352.175$. For low $n$, the states are similar to discretized harmonic oscillator wavefunctions, but for higher energies the states become localized away from the center of the parabolic potential.\label{fig:estates}}
 \end{figure}

The eigenstates are $\Psi(\vec{x})=\Psi_{n_x}(x)\Psi_{n_y}(y)\Psi_{n_z}(z)$, where the sum $\Psi_{n_x}(x)=\sum_{j} f_{j}^{n_x}w_j(x)$ runs over one direction of the lattice.  The site-dependent weights are \cite{rey:2005},
\begin{equation}
f_{j}^{n} = \left\{ \begin{array}{lr} \frac{1}{\pi} \int_{0}^{2\pi} dx~ce_{n}(x,-\alpha) \cos(2jx), & \mbox{$n$ even} \\ \frac{1}{\pi} \int_{0}^{2\pi} dx~se_{n+1}(x,-\alpha) \sin(2jx), & \mbox{$n$ odd} \end{array} \right.,
\end{equation}
where $ce$ and $se$ are the even and odd periodic $\pi$ solutions of the Mathieu equations with parameter $\alpha$. The probability distributions along one lattice direction for several energy levels are plotted in Figure \ref{fig:estates}.  The low energy states are similar to discretized harmonic energy eigenstates with an oscillator frequency $\omega \sqrt{m/m^{*}}$, where $m^{*}=\frac{\hbar^2}{2d^2J}$ \cite{zwerger:2003} is the effective mass resulting from the lattice dispersion relation (Eq. \ref{eqn:dispersion}).  As the energies exceed the bandwidth $4J$ of the uniform lattice potential, the states become localized to a few lattice sites; the localized states are found farther from the center of the harmonic trap as $n$ increases.  The localized states were observed to affect transport \cite{ott:2004,strohmaier:2007} and were detected directly in Ref. \cite{ott:2004}.  The energies of these states are inside the uniform lattice band-gap, reminiscent of surface states in solids. 

To calculate thermodynamic quantities using these states we directly calculated the grand canonical potential for bosons by summing over all states in 3D \cite{pathria:1996},
\begin{equation}
\Omega = -\beta^{-1} \sum_{n_x,n_y,n_z} \log \left(1-e^{\beta E_{n_x,n_y,n_z}} \mathfrak{z}^{-1} \right), \label{eqn:grandcanonical}
\end{equation}
where $\beta=1/k_B T$ and $\mathfrak{z}=e^{\beta \mu}$ is the fugacity  ($\mu$ is the chemical potential and $k_B$ is Boltzmann's constant). The number of particles,
\begin{equation}
N = \sum_{n_x,n_y,n_z} \frac{1}{e^{\beta E_{n_x,n_y,n_z}} \mathfrak{z}^{-1} - 1}, \label{eqn:n}
\end{equation}
and energy,
\begin{equation}
U = \sum_{n_x,n_y,n_z} \frac{E_{n_x,n_y,n_z}}{e^{\beta E_{n_x,n_y,n_z}} \mathfrak{z}^{-1} - 1}, \label{eqn:u}
\end{equation}
were used to calculate entropy,
\begin{equation}
S = -\frac{\Omega}{T} + \frac{U}{T} - \frac{\mu N}{T}.
\end{equation}
We also used Eq. \ref{eqn:grandcanonical}, \ref{eqn:n}, and \ref{eqn:u} to calculate the thermodynamics for the harmonic trap by using the harmonic trap eigenenergies $E_{n_x,n_y,n_z}=\hbar \omega (n_x + n_y + n_z)$ \cite{ketterle:1996}. 

To calculate the final temperature and condensate fraction in the lattice based on adiabatically loading atoms from the harmonic trap (at a given initial temperature and number of atoms), we first calculated the entropy and fugacity of the atoms in the harmonic trap. Then we varied temperature and fugacity so that the final entropy and number were identical in the combined parabolic--lattice potential.  

\section{Quasimomentum Distribution \label{sect2}}

In this section, we determine the quasimomentum distribution for non-interacting particles in the joint parabolic--lattice potential.   For particles in a periodic potential, the quasimomentum operator plays the role of the momentum operator for free particles.  The Hamiltonian from Eq. \ref{eqn:ham2} for atoms in a lattice--parabolic potential written in terms of the quasimomentum and position operators is
\begin{equation}
H = 2J \sum_{i=x,y,z} \left[1-\cos \left( \frac{d}{\hbar} \hat{q}_i \right) \right] + \frac{1}{2} m \omega^2 \sum_{i=x,y,z} \hat{x}_i^2. \label{eqn:ham3}
\end{equation}

To calculate the quasimomentum distribution, the single-particle eigenstates must be transformed into functions of quasimomentum \cite{rey:2005}: $\tilde{\Psi}_{n_x,n_y,n_z}(\vec{q}) = f_{n_x}(q_x)f_{n_y}(q_y)f_{n_z}(q_z)$ (see Appendix \ref{sec:appendix}), where
\begin{equation}
f_{n_x}(q_x) = \left \{ \begin{array} {lr} \frac{(-1)^{n_x/2}}{\sqrt{\pi}} ce_{n_x} \left[ \frac{\pi}{2} \left(1- \frac{q_x}{q_B}\right), \alpha \right], & \mbox{$n$ even} \\ \frac{(-1)^{(1-n_x)/2}}{\sqrt{\pi}} se_{n_x+1} \left[ \frac{\pi}{2} \left(1- \frac{q_x}{q_B}\right), \alpha \right], & \mbox{$n$ odd} \end{array} \right.;
\end{equation}
analogous equations apply for $f_{n_y} (q_y)$ and $f_{n_z} (q_z)$. For the finite temperature, statistical distribution of the Hamiltonian in Eq. \ref{eqn:ham3} we add a Bose-Einstein factor,
\begin{equation}
\rho(\vec{q}) = \sum_{n_x,n_y,n_z} \left| \tilde{\Psi}_{n_x,n_y,n_zq}(\vec{q}) \right|^2 \frac{1}{e^{\beta E_{n_x,n_y,n_z}} \mathfrak{z}^{-1} -1 }, \label{eqn:quasidist}
\end{equation}
where the sum is over all eigenstates.  Fitting images of the quasimomentum distribution to $\rho(\vec{q})$ calculated using Eq. \ref{eqn:quasidist} is infeasible given modest computational resources.  Fortunately, by writing the Hamiltonian in terms of quasimomentum, we are justified in using the semi-classical distribution, since the rapid spatial variations in the potential and wavefunction have disappeared \cite{gerbier:2007b,sundaram:1999}.  The semi-classical approximation is appropriate in this system because the quasimomentum and position operators are conjugate \cite{pitaevskii:1980}
\begin{equation}
\left[\hat{q},\hat{x} \right] = -i\hbar.
\end{equation}

In the semi-classical distribution, the operators in the Hamiltonian are replaced with classical variables, and each particle occupies the minimum uncertainty volume in phase space consistent with the commutation relation. Using this approximation, the three-dimensional, finite-temperature quasimomentum distribution for bosons in a lattice--parabolic trap is (where $\hbar k=q$),
\begin{widetext}
\begin{equation}
\rho(k_1,k_2,k_3) = \frac{1}{(2\pi)^3} \int \int \int dx dy dz~\frac{1}{\exp \left\{ 2J\beta \sum_{i=x,y,z} \left[1-\cos(k_i d) \right] + \frac{1}{2}\beta m \omega^2 r^2 \right\} \mathfrak{z}^{-1} - 1}. \label{eqn:quasidist2}
\end{equation}

After integrating over the spatial degrees of freedom,
\begin{equation}
\rho(k_1,k_2,k_3) = \frac{\sqrt{\pi}}{8\pi^2} \left(\frac{\beta m \omega^2}{2} \right)^{3/2} Li_{3/2} \left\{ \mathfrak{z} e^{-2J\beta \sum_{i=x,y,z} \left[1-\cos(k_i d)\right] } \right\},
\end{equation}
\end{widetext}
where $Li_n(u) = \sum_{k=1}^{\infty} u^k/k^n$ is the polylogarithm function.  This equation reveals that, in the thermal limit ($\mathfrak{z} \ll 1$), the quasimomentum distribution is solely determined by the dimensionless parameter $1/J\beta=k_{B} T/J$, which characterizes the ratio of the temperature to the band width.  In the quantum degenerate regime ($\mathfrak{z} \approx 1$), the fugacity and the parameter  $k_B T/J$ both control the shape of the quasimomentum distribution.

To image the quasimomentum distribution, we employ ``bandmapping'' followed by standard time-of-flight (TOF) absorption imaging.  Bandmapping involves turning off the lattice potential slowly with respect to the bandgap, but quickly with respect to the slowly varying harmonic trapping potential.  This procedure maps a state of quasimomentum, $q$, to a state of momentum $q$.  This technique has been demonstrated by several groups \cite{greiner:2001,denschlag:2002,kohl:2005b}, but we will show in Sec. IV that it has limits of validity not previously discussed in the literature.

Each pixel in the two-dimensional absorption image is a column integral of the quasimomentum distribution along the imaging axis.   We use an atypical imaging geometry in our experiment: the imaging axis makes $90^{\circ}$ and $\pm45^{\circ}$ angles with the lattice directions, which define the quasimomentum axes in space after TOF.  For our experiment, the coordinates  $\alpha_1$ and $\alpha_2$ define the imaging plane; the integration direction is along coordinate $\alpha_3$; and $k_1$,  $k_2$, and $k_3$ are coordinates along the lattice axes given by,
\begin{eqnarray}
k_1 & = & \frac{1}{\sqrt{2}} \alpha_1 - \frac{1}{\sqrt{2}} \alpha_2, \\
k_2 & = & \frac{1}{2} \alpha_1 + \frac{1}{2} \alpha_2 + \frac{1}{\sqrt{2}} \alpha_3, \\
k_3 & = & \frac{1}{2} \alpha_1 + \frac{1}{2} \alpha_2 - \frac{1}{\sqrt{2}} \alpha_3.
\end{eqnarray}

The two-dimensional distribution integrated along the direction of the probe beam is
\begin{equation}
\rho(\alpha_1,\alpha_2) = \int d\alpha_3~\rho(k_1,k_2,k_3),
\end{equation}
which can be expressed as a doubly infinite sum
\begin{widetext}
\begin{equation}
\rho(\alpha_1,\alpha_2) = A \sum_{n=1}^{\infty} \frac{e^{-2nJ\beta \left[3-\cos(b)\right]} \mathfrak{z}^{n}}{n^{3/2}} \left\{ 2(\pi-a) I_o\left[\cos(a)2nJ\beta\right] + \sum_{j=1}^{\infty} I_j \left[\cos(a)2nJ\beta\right] 4 \frac{\sin\left[j(\pi-a)\right]}{j} \right\}, \label{eqn:quasifit}
\end{equation}
\end{widetext}
where $A$ is a proportionality constant, $a = |\alpha_1+\alpha_2|/\sqrt{2}$, $b=|\alpha_1-\alpha_2|/\sqrt{2}$ and $I_j(x)$ is the modified Bessel function of the first kind.

\begin{figure}
 \pic{0.5}{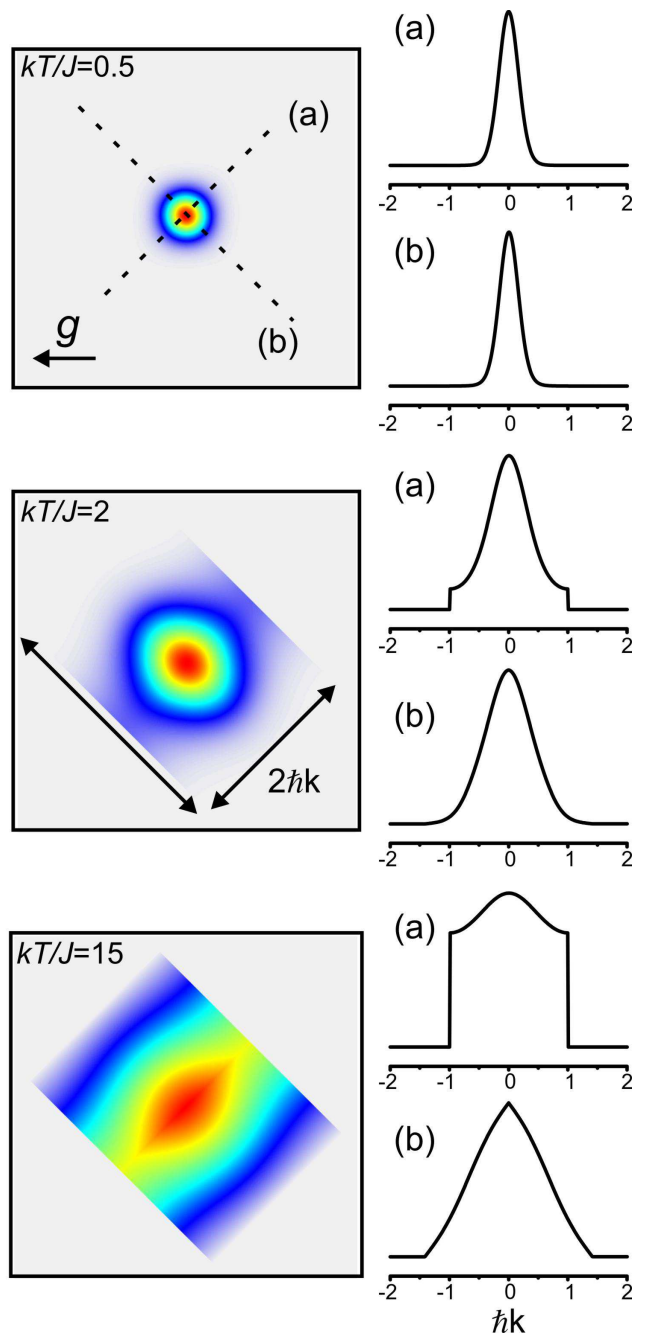}
 \caption{(Color Online) Examples of predicted images from Eq. 26 for $k_B T/J=0.5$, $k_B T/J=2$ and $k_B T/J=15$ (in the thermal limit).  Cross-sections through the distribution are also shown. Along cross-section A, the distribution has a sharp edge at the Brillouin-zone momentum at high $k_B T/J$. For low values of $k_B T/J$, this edge vanishes and the distribution is similar to a momentum distribution for atoms confined in a harmonic trap.  The direction of gravitational acceleration, $g$, relative to the imaging plane for our experiment is labeled by an arrow.\label{fig:quasidists}}
 \end{figure}

Examples of predicted images for $\mathfrak{z}=0.5$ and for different values of $k_B T/J$ are shown in Figure \ref{fig:quasidists}. A characteristic feature of these distributions is the sharp edge at the Brillouin zone momentum for high $k_B T/J$.  We have verified that distributions calculated using a direct sum over eigenstates (Eq. \ref{eqn:quasidist}) agree with the semi-classical result for the range of $k_B T/J$ used in this paper.

\section{Measuring Temperature using Quasimomentum Distributions \label{sect3}}

Images of the quasimomentum distribution obtained for a range of temperatures and lattice depths are fit to the semi-classical result (Eq. \ref{eqn:quasifit}).  In this section, we discuss how the fit is used to determine $k_B T/J$, which is compared with thermodynamic predictions for non-interacting particles.  All data in this section and Sec. \ref{sect5} are taken above the critical temperature for Bose-Einstein condensation.

The procedure by which we prepare ultra-cold gases of $^{87}$Rb atoms confined in a lattice--parabolic potential has changed somewhat compared with Refs. \cite{white:2008,mckay:2008}. $^{87}$Rb atoms in the $F=1$, $m_F=-1$ state are evaporatively cooled in a two-stage process.   The atom gas is first cooled using forced radio-frequency evaporation in a magnetic quadrupole trap with a 300~G/cm gradient along the symmetry axis.  Before atom loss induced by Marojana transitions becomes significant, the atom gas is transferred into a hybrid magnetic--optical trap by slowly turning on a 7~W, 1064~nm Gaussian laser beam focused to a 90~$\mu$m waist; the magnetic field gradient is simultaneously reduced to produce a force on the atoms equal to gravity. The second stage of evaporative cooling progresses by increasing the quadrupole gradient and reducing the optical power \cite{hung:2008}.  Finally, the magnetic gradient is again relaxed to balance gravity and the dipole beam power is reduced to 0.4~W, which creates a harmonic trap with oscillator frequencies 56.4 $\pm$ 0.7, 29.2 $\pm$ 1.5, and 39.1 $\pm$ 0.2 Hz.  The temperature $T_{ho}$ of the atom gas in this trap is controlled by altering the evaporative cooling sequence, both by changing the final radio-frequency applied during cooling in the quadrupole trap and the minimum optical power used during cooling in the hybrid trap.

After the dipole beam power is decreased to 0.4~W, the optical lattice beams ($\lambda=812$~nm) are super-imposed on the hybrid trap (see Ref. \cite{mckay:2008}  for more information on the lattice geometry).  Bandmapping is implemented by turning off the optical lattice beams in 750~$\mu$s using a linear ramp of the optical lattice power.  Following bandmapping, the 1064~nm beam and the quadrupole magnetic trap are turned off and an absorption image is taken after 8--20 ms of time-of-flight (TOF).

\begin{figure}
 \pic{0.5}{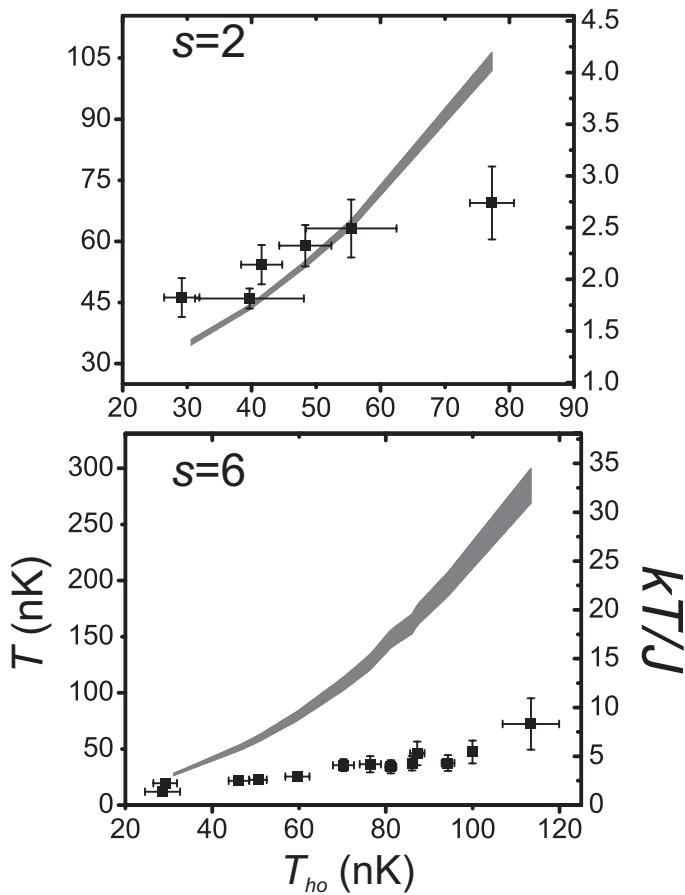}
 \caption{Measured temperature $T$ in the lattice obtained by fitting the quasimomentum distribution for $s=2$ (top) and $s=6$ (bottom) as the temperature $T_{ho}$ in the harmonic trap is varied.  The vertical error bars are determined by the standard deviation in several measurements that are averaged for each point and a 10\% uncertainty in the Brilluoin-zone width after time-of-flight.  The error bars in $T_{ho}$ represent statistical uncertainty. There is also a 5\% systematic uncertainty in the TOF, which is not included in the error bars.   The curve is the thermodynamic prediction for isentropic transfer into the lattice and non-interacting particles; the width of the theory curve reflects error in $N$ and $\omega$.\label{fig:tmeas}}
 \end{figure}

Figure \ref{fig:tmeas} shows the measured temperature $T$ in the lattice as the temperature in the harmonic trap is varied.  Images taken after bandmapping and time-of-flight expansion are fit to Eq. \ref{eqn:quasifit} using a non-linear least squares solver, with $A$, $k_B T/J$, the maximum optical depth, and fugacity as free parameters.  Data are shown for $s=2$ and $6$. While Eqs. \ref{eqn:ham2} and \ref{eqn:ham3} require corrections for next-nearest-neighbor tunneling at $s=2$, the next-nearest-neighbor tunneling energy is $14\%$ of $J$, which results only in a small deviation from the tight-binding dispersion relation.  The fitted value of $k_B T/J$ is converted into temperature $T$ in the lattice using a truncated Fourier series band calculation for $J$, based on a calibration of $s$ within 5\% using Raman-Nath diffraction \cite{morsch:2006}.  The number of atoms $N$ ranged from $3\times10^{3}$--$1.4\times10^{5}$  and the fitted fugacity from 0.35--0.74 for the data shown in Fig. \ref{fig:tmeas}.

The measured temperature in the lattice agrees with the prediction for isentropic transfer---shown by the curve in Fig. \ref{fig:tmeas}---when $T_{ho}$ leads to a thermodynamically predicted $k_B T/J \lesssim 2.5$.  The theoretical prediction is calculated using the measured $T_{ho}$ and $N$, as discussed in Sec. \ref{sect1}.  We also include the change in the parabolic confinement induced by the Gaussian profile of the lattice laser beams.   We assume that the overall confinement is spherically symmetric, with a harmonic oscillator frequency
\begin{equation}
\omega = \sqrt{\omega_0^2 + \frac{8sE_R}{w^2 m}}, \label{eqn:omega}
\end{equation}
where $\omega_0=40$Hz is the geometric mean of the oscillator frequencies for the hybrid trap, and $w=120\pm 10 \mu m$ is the waist of the optical lattice beams.

For higher $T_{ho}$, the measured temperature in the lattice is systematically lower than the thermodynamic prediction.  Effects due to interactions can be straightforwardly ruled out as the source of this discrepancy as the ratio of mean energy to estimated interaction energy per particle is greater than 30 for the data in Fig. \ref{fig:tmeas}. Also, a failure of adiabaticity can be ruled out as the measured temperature is less than the thermodynamic expectation.  Furthermore, the temperature in the lattice measured using an alternate technique agrees with the thermodynamic prediction, as demonstrated in Sec. \ref{sect5}.  In the next section, we show that the deviation is consistent with a failure of bandmapping for states of high quasimomentum, which are populated at high $k_B T/J$.

\section{Simulation of Bandmapping in 1D \label{sect4}}

To investigate the validity of bandmapping as a technique for measuring the quasimomentum distribution, we numerically simulate the process using a Crank-Nicolson solver \cite{thijssen:2007}.  We calculated the time evolution of the single-particle eigenstates of the Hamiltonian in Eq. \ref{eqn:ham1} in 1D, with the lattice depth $s$ given by:
\begin{equation}
s(t) = \left\{ \begin{array} {lr} 6, & t<0 \\ 6\left(1-\frac{t}{750 \mu s}\right), & 750\mu s \ge t \ge 0 \end{array} \right.,
\end{equation}
and the corresponding harmonic oscillator frequency determined by Eq. \ref{eqn:omega}.  The Wannier function at each lattice site was approximated by the ground state wavefunction of a harmonic oscillator with angular frequency $2E_R \sqrt{s}/\hbar$ \cite{zwerger:2003}.  A total of 240 lattice sites were included in the simulation up to $t= 750 \mu s$, when the bandmapping process is completed (i.e., $s=0$). The wavefunctions were then propagated for $20$ms of free-evolution using the kernel,
\begin{equation}
g(x^{\prime}, x, t) = \sqrt{\frac{m}{2\pi i \hbar t}} e^{i m (x-x^{\prime})^2/2\hbar t},
\end{equation}
which was convolved with the wavefunction obtained from the solver.  The wavefunctions $\Psi_n^{\prime}(x^{\prime})$ obtained via this procedure are the single particle states when an image is taken.

\begin{figure}
 \pic{0.5}{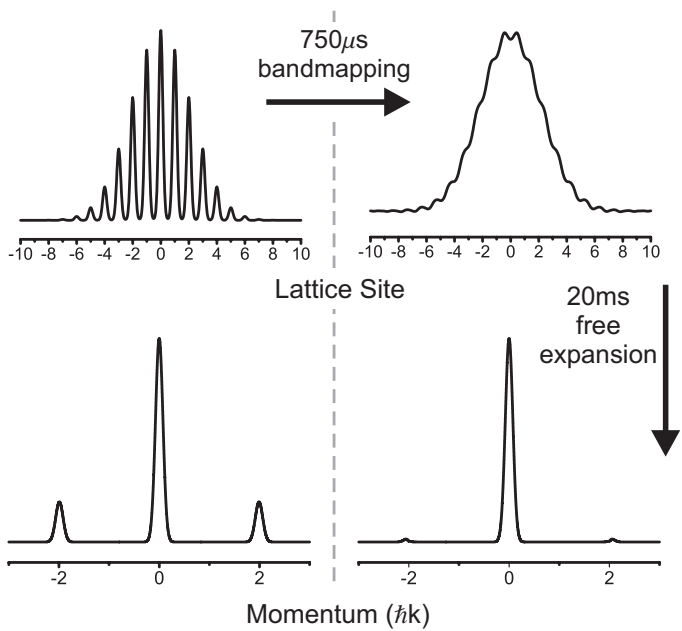}
 \caption{Numerical simulation of bandmapping for the $n=0$ state.  The single-particle eigenstate is shown at the top left; the wavefunction after 20 ms of TOF, revealing the corresponding momentum distribution, is displayed at the bottom left.  The wavefunction after bandmapping is displayed at the top right, with the wavefunction after 20 ms TOF shown at the bottom right.\label{fig:bandmapsim}}
 \end{figure}

Figure \ref{fig:bandmapsim} shows the result of this simulation for the $n=0$ eigenstate.  The effect of bandmapping is to change the spatial wavefunction so that the momentum distribution after bandmapping is the quasimomentum distribution before bandmapping. Since the quasimomentum wavefunction is given by the Fourier series of the coefficients of the Wannier functions (see Eq. \ref{eqn:quasi2} in Appendix \ref{sec:appendix}), and momentum is the Fourier transform of the wavefunction, bandmapping is equivalent to removing the modulation of the wavefunction at the lattice spacing. Standard TOF imaging, after sufficient expansion time, therefore reveals the quasimomentum distribution.

\begin{figure}
 \pic{0.5}{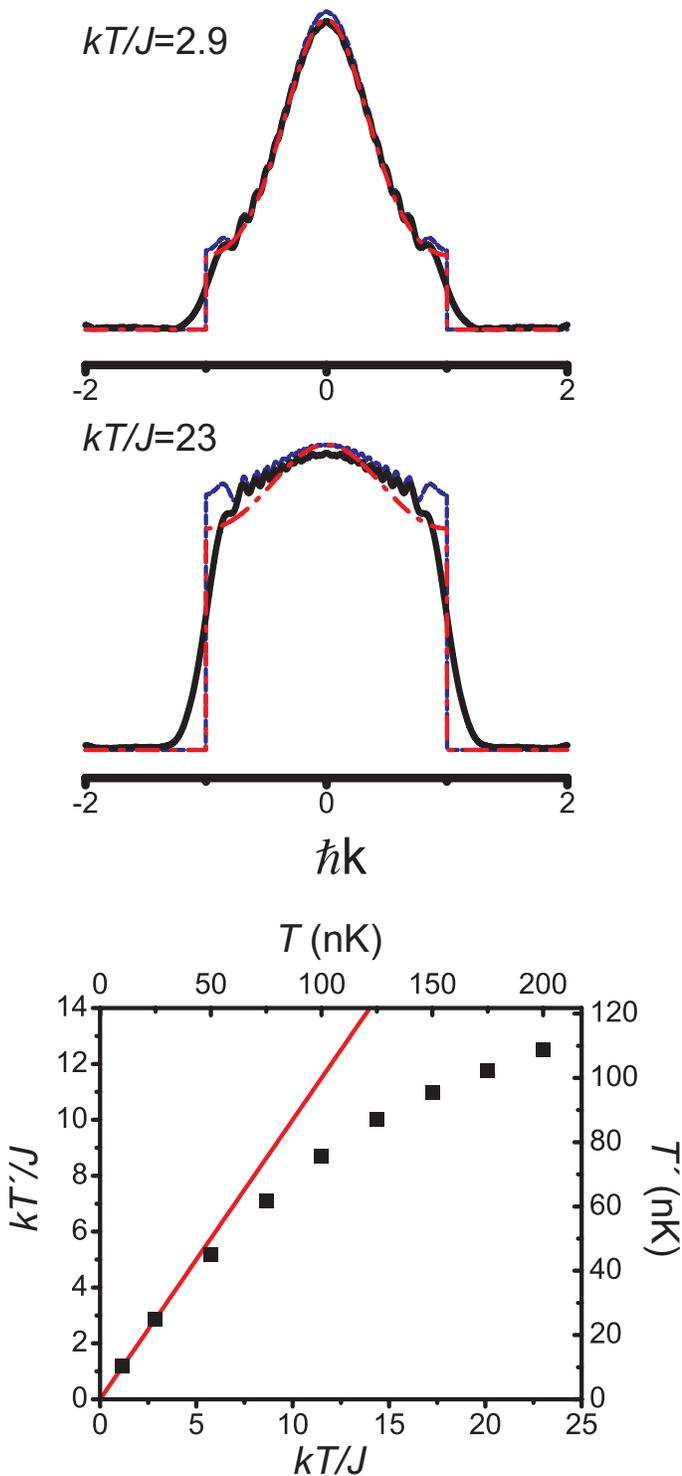}
 \caption{(Color Online) Failure of bandmapping for high $k_B T/J$. The top panels show samples of the finite temperature bandmapped distribution (solid black line), the finite-temperature quasimomentum distribution (dashed blue line), and the semi-classical fit to the bandmapped distribution (dashed and dotted red line). The bottom panel shows the temperature $T^{\prime}$ obtained using a semi-classical fit to a 1D simulation of imaging at temperature $T$ after bandmapping and TOF.  The red line has a slope of 1 and is present to guide the eye. \label{fig:tmeassim}}
 \end{figure}

To compare with the results of Sec. \ref{sect3}, we use the results of this calculation to create simulated images in the thermal limit for a range of $k_BT/J$.   We restrict the calculation to 1D and simulate the time-evolution of the 150 lowest energy eigenstates.  The predicted image is calculated by summing the eigenstate probability distributions after bandmapping and time-of-flight, weighted by appropriate Boltzmann factors ($\mathfrak{z} \ll 1$):
\begin{equation}
\rho(x^{\prime}) = \frac{\mathfrak{z}}{N} \sum_{n} e^{-\beta E_n} \left|\Psi_n^{\prime}(x^{\prime})\right|^2.
\end{equation}
Predicted images simulated using this scheme are shown as solid black lines in Fig. \ref{fig:tmeassim} for $k_B T/J=2.88$ and $23$; the coordinate $x^{\prime}$ after TOF is converted to quasimomentum in the lattice using $q=m x^{\prime}/\tau$, where $\tau$ is the free evolution time.  The predicted images are compared with the exact 1D quasimomentum distribution $\rho(q) = \frac{\mathfrak{z}}{N}\sum_n e^{-\beta E_n} \left|\tilde{\Psi}_n(q) \right|^2$ calculated in thermal limit (dashed blue line).

Apparent in Fig. \ref{fig:tmeassim} is a failure of bandmapping at high quasimomentum.  During bandmapping, the bandgap at the Brilluoin zone edge shrinks, and the adiabatic timescale for changes in the lattice depth at high $q$ is extended.  Atoms with high quasimomentum therefore make diabatic transitions out of the Brillouin zone during bandmapping, leading to significant smoothing of the sharp edge in the image at $q=\pi/d$ for high $k_B T/J$. Because of this effect, images taken after bandmapping and time-of-flight at high $k_B T/J$ are not images of the quasimomentum distribution. We find that in this simulation that we are not very sensitive to the choice of bandmapping time. Indeed, in the literature, bandmapping times of 20~$\mu$s\cite{denschlag:2002}, 200~$\mu$s\cite{schneider:2008,mckay:2008,white:2008}, 1~ms\cite{kohl:2005b} and 2~ms\cite{greiner:2001} have been employed.

To understand the impact of this problem on the method employed in Sec. \ref{sect3}, simulated images for $k_B T/J=$1--23 are fit to the 1D semi-classical quasimomentum distribution
\begin{equation}
\rho(q) = A e^{-2\beta^{\prime} J \left[1-\cos(\pi q d) \right]},
\end{equation}
where $A$ and $\beta^{\prime}$ are free parameters. The top panel of Fig. \ref{fig:tmeassim} shows the fitted $k_B T^{\prime}/J$ as $k_B T/J$ (used to create the distribution) is varied.   The fit systematically underestimates the temperature for $k_B T/J \gtrsim 3$, implying that a failure of bandmapping is responsible for the discrepancy at high $k_B T/J$ evident in Fig. \ref{fig:tmeas} between the thermodynamic prediction and the fitted temperature.  We are unable to make direct, quantitative comparisons between simulations and our data because of the computational complexity of calculating a 3D density distribution after bandmapping, TOF, and projection onto the imaging plane. 

\section{Measuring Temperature using In-Trap Size \label{sect5}}

Because bandmapping fails to produce quasimomentum distributions at high temperature, we have explored another method for thermometry.  We determine temperature using the measured size of the gas while it is confined in the parabolic--lattice potential.  This technique not only overcomes the limitations of bandmapping, but also has sensitivity superior to fitting the quasimomentum distribution at high $k_B T/J$.  The quasimomentum distribution is largely insensitive to temperature once states at the edge of the Brilluoin zone are significantly populated.  In contrast, the density profile always depends strongly on temperature, since the localized states---while nearly uniform in quasimomentum---extend to larger radii in the parabolic potential as their energy increases.

\begin{figure}
\pic{0.5}{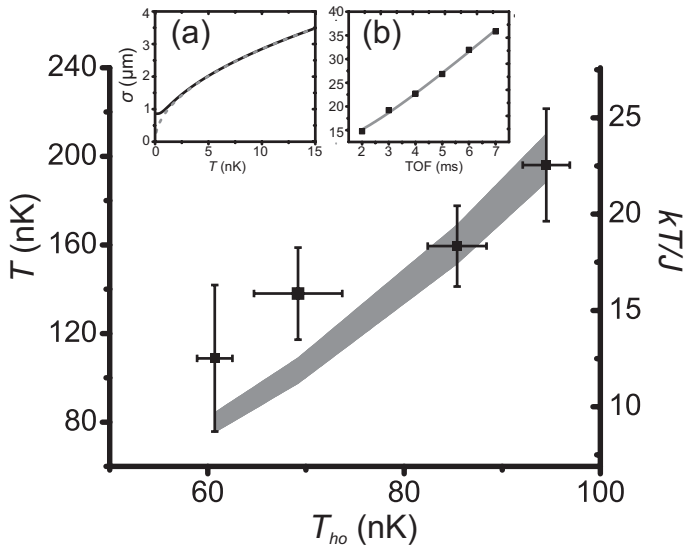}
 \caption{Temperature measured using the in-trap size for $s=6$.  The curve is the theory prediction assuming adiabatic transfer into the lattice; the width of this curve is determined by uncertainty in $N$ and $\omega$.  The error bars in $T$ reflect the uncertainty in the extrapolation to the in-trap size, and the error bars in $T_{ho}$ represent the uncertainty in using standard TOF expansion to measure the temperature of the gas before loading into the lattice. Inset (a) shows the exact calculation of the width in 1D (black line) from Eq. \ref{eqn:widthexact} versus the semiclassical value $\sqrt{kT/m\omega^2}$ (dashed grey line). The agreement is exact except at very low temperatures. Inset (b) shows a sample set of expansion data used to determine the in-trap size. The line is a fit to the data of the form $\sigma=\sqrt{\sigma_0^2+A^2 t^2}$ where $A$ and $\sigma_0$ are fit parameters.  \label{fig:tmeasintrap}}
 \end{figure}

The relation between in-trap size and temperature is given by the exact expression for the mean-squared width of the cloud in the lattice along a direction $\hat{r}$,
\begin{eqnarray}
\langle |\hat{r}\cdot \vec{x}|^2 \rangle & = & \frac{\mathfrak{z}}{N} \sum_{n_x,n_y,n_z} \int d^{3}x~|\hat{r}\cdot\vec{x}|^2 \ldots \nonumber \\
& & \ldots  |\Psi_{n_x,n_y,n_z}(\vec{x})|^2 e^{-\beta E_{n_x,ny,n_z}}. \label{eqn:widthexact}
\end{eqnarray}
However, we can use the semiclassical expression, Eq. \ref{eqn:quasidist2}, for the width ($\mathfrak{z} \ll 1$),
\begin{equation}
\langle |\hat{r}\cdot \vec{x}|^2 \rangle = \frac{\int d^{3}x~|\hat{r}\cdot\vec{x}|^2 e^{-m\omega^2 x^2/2k_B T}}{\int d\vec{x} e^{-m\omega^2 x^2/2k_B T}}, \label{eqn:widthsemi}
\end{equation}
which gives the familiar expression $\langle |\hat{r}\cdot \vec{x}|^2 \rangle = k_B T/m\omega^2$. Comparing Eq. \ref{eqn:widthsemi} to the width numerically calculated from Eq. \ref{eqn:widthexact} (Fig. \ref{fig:tmeasintrap}a), the agreement is excellent as long as the temperature is greater than the spacing between the first two eigenstates given by Eq. \ref{eqn:eigenergies}.

We cannot directly measure the in-trap density distribution using our apparatus.  Therefore, we measure the density profile after snapping off the lattice and relatively short (i.e., less than 10 ms) TOF; the size of the gas in the trap is inferred from extrapolation.  We determine the r.m.s. size of the gas by fitting images taken after expansion time $t$ to a Gaussian profile.  We then extrapolate to the r.m.s.\ size in the trap by assuming for short expansion times that the gas expands with a momentum distribution similar to that of a gas confined in a harmonic potential.  We therefore fit the r.m.s.\ size after expansion time $t$ to $\sigma(t)=\sqrt{\sigma_0^2+At^2}$, leaving $\sigma_0$ and $A$ as free parameters \cite{bruun:2000}.    The temperature of the gas is inferred from the fitted value of $\sigma_0$ and the measured value of $\omega$ (Fig. \ref{fig:tmeasintrap}b).

Figure \ref{fig:tmeasintrap} shows the measured temperature in the lattice for $s=6$ as the temperature in the harmonic trap is varied. The agreement with the thermodynamic prediction (grey curve) is excellent, implying that the disagreement with this prediction evident in Fig. \ref{fig:tmeas}b is due to the failure of bandmapping at high quasimomentum.  

\section*{Conclusions}

In conclusion, we have determined the absolute temperature of atoms confined in a lattice in the thermal limit by employing two methods: fitting quasimomentum distributions obtained via bandmapping and measuring the in-trap size of the gas.  These methods may be useful for verifying other methods of thermometry and can be applied to fermionic gases.  Furthermore, determining temperature in the thermal limit will prove useful for future studies of unexplained transport phenomena in optical lattices at relatively high temperature \cite{mckay:2008,ferlaino:2002}.  Through numerical simulation, we have also demonstrated that bandmapping fails to produce accurate quasimomentum distributions at high temperature (or when high quasimomentum states are occupied).   We expect this discovery to have an important impact on this technique, which has been applied in many optical lattice experiments \cite{denschlag:2002,schneider:2008,mckay:2008,white:2008,kohl:2005b,greiner:2001}.

\appendix*

\section{Quasimomentum Transformations \label{sec:appendix}}

The quasimomentum basis states, $\Phi_{\vec{q}}(\vec{x})$, are a complete set, so any wavefunction can be written in the form (considering only a single band),
\begin{equation}
\Psi(\vec{x}) = \int\int \int_{-q_{B}}^{q_{B}} d^{3}q~f(\vec{q}) \phi_{\vec{q}}(\vec{x}). \label{eqn:quasi1}
\end{equation}
The function $f(\vec{q})$ is the quasimomentum-space wavefunction. The transformation from spatial to quasimomentum wavefunction is,
\begin{equation}
f(\vec{q}) = \int d^3x~\phi^{*}_{\vec{q}}(\vec{x}) \Psi(\vec{x}).
\end{equation}
In particular, if we have a wavefunction defined in terms of the Wannier functions,
\begin{equation}
\Psi(\vec{x}) = \sum_{j} a_j w_j(\vec{x}), 
\end{equation}
where $j$ labels lattice sites, then, 
\begin{eqnarray}
f(\vec{q}) & = & \int d^{3}x~\phi^{*}_{\vec{q}}(\vec{x}) \sum_{j} a_j w_j(\vec{x}), \nonumber \\
& = & \int d^{3}x~\phi^{*}_{\vec{q}}(\vec{x}) \sum_{j} a_j \int d^{3}q^{\prime}~e^{i \vec{q^{\prime}}\cdot \vec{R}_j/\hbar} \phi_{\vec{q^{\prime}}}(\vec{x}), \nonumber \\
& = &  \sum_{j} a_j \int d^{3}q^{\prime}~e^{i \vec{q^{\prime}}\cdot \vec{R}_j/\hbar} \int d^{3}x~ \phi^{*}_{\vec{q}}(\vec{x}) \phi_{\vec{q^{\prime}}}(\vec{x}), \nonumber \\
& = & \sum_{j} a_j \int d^{3}q^{\prime}~e^{i \vec{q^{\prime}}\cdot \vec{R}_j/\hbar} \delta^{3}(\vec{q}-\vec{q^{\prime}}), \nonumber \\
& = & \sum_{j} a_j e^{i \vec{q}\cdot \vec{R}_j/\hbar}. \label{eqn:quasi2}
\end{eqnarray}

Therefore, the quasimomentum-space wavefunction is given by the Fourier series of the Wannier function coefficients. Finally, the transformation between the quasimomentum wavefunction and momentum (taking the Fourier transform of Eq. \ref{eqn:quasi1}),
\begin{equation}
\tilde{\Psi}(\vec{p}) = \int \int \int_{-q_B}^{q_B} d^{3}q~f(\vec{q}) \tilde{\Phi}_{\vec{q}}(\vec{p}).
\end{equation}

The momentum-space Bloch wavefunctions are
\begin{equation}
\tilde{\Phi}_{\vec{q}}(\vec{p}) = \tilde{w}_{0}(\vec{p}) \sum_{j} \delta^{3}(\vec{p}-\vec{q}-\vec{P}_j),
\end{equation}
where $\tilde{w}_{0}(\vec{p})$ is the Fourier transform of the Wannier function at site $\vec{R}=0$ and $\vec{P}_j$ is a reciprocal lattice vector. Therefore,
\begin{equation}
\tilde{\Psi}(\vec{p}) = \int \int \int_{-q_B}^{q_B} d^{3}q~f(\vec{q}) \tilde{w}_{0}(\vec{p}) \sum_{j} \delta^{3}(\vec{p}-\vec{q}-\vec{P}_j). 
\end{equation}

For $|\vec{p}|<q_B$ the only way for the integrand to be non-zero is if $\vec{P}_j=0$, since all other reciprocal lattice vectors lie outside the Brillouin zone, so,
\begin{equation}
\tilde{\Psi}(\vec{p}) = f(\vec{p}) \tilde{w}_{0}(\vec{p}).
\end{equation}

\begin{acknowledgments}
We acknowledge funding from the National Science Foundation (award 0448354), the Army Research Office (W911NF-08-1-0021), the DARPA OLE program, and the Sloan Foundation.  D. McKay acknowledges support from NSERC.
\end{acknowledgments}

\bibliography{latticethermo}

\end{document}